\documentstyle[aps,epsfig,multicol,amsmath,amssymb]{revtex}
\begin{document}
\title{Statistics of the critical percolation backbone\\
with spatial long-range correlations}

\author{A. D. Ara\'ujo$^{1,2}$, A. A. Moreira$^{1}$, R. N. Costa
Filho$^{1}$, and J. S. Andrade Jr.$^{1}$}
\address{$^1$Departamento de F\'{\i}sica, Universidade Federal do
Cear\'a, 60451-970 Fortaleza, Cear\'a, Brazil.\\$^2$Departamento de
F\'{\i}sica, Universidade Estadual Vale do Acara\'u, 62040-370 Sobral, Cear\'a,
Brazil.\\}

\date{\today}

\maketitle

\begin{abstract}
We study the statistics of the backbone cluster between two sites
separated by distance $r$ in two-dimensional percolation networks
subjected to spatial long-range correlations. We find that the
distribution of backbone mass follows the scaling {\it ansatz},
$P(M_B)\sim M_B^{-(\alpha+1)}f(M_B/M_0)$, where $f(x)=(\alpha+
\eta x^{\eta}) \exp(-x^{\eta})$ is a cutoff function, and $M_0$ 
and $\eta$ are cutoff parameters. Our results from extensive
computational simulations indicate that this scaling form is
applicable to both correlated and uncorrelated cases. We show that the
exponent $\alpha$ can be directly related to the fractal dimension
of the backbone $d_B$, and should therefore depend on the imposed
degree of long-range correlations.
\end{abstract} 

\draft

\begin{multicols}{2}

\section{Introduction}

Percolation is a useful model to study a variety of systems in many
fields of science displaying structural disorder and statistical
self-similarity. In particular, the percolation geometry has been
frequently used as a simple paradigm to investigate fluid flow in
porous media \cite{Bear72,Dullien79,Sahimi95}. In this type of
problem, the geometry underlying the system can be very complex and
display heterogeneous features over a very wide range of length
scales, going from centimeters to kilometers \cite{King90}. For
example, an open question in the modeling process of oil recovery is
the effect of the connectedness of the porous medium on the dynamical
process of fluid displacement. If the pore space is so poorly
connected as to be considered macroscopically heterogeneous, one
expects the overall behavior of the flowing system to display
significant anomalies. In this way, it is important to investigate the
physics of disordered media at a marginal state of connectivity, for
example, in terms of the geometry of the spanning cluster at the
percolation threshold \cite{Stauffer94,Havlin96}.

The most relevant subset of the percolation cluster for transport is
the conducting backbone. It can be defined as the cluster that carries
the current when a voltage difference is applied between two
sites. Thus the backbone structure alone determines the conductivity
of the whole percolation network. Recently, Barth\'el\'emy {\it et
al.} \cite{Barth99} have shown that the average mass of the backbone
$\langle M_B \rangle$ connecting two sites in a two-dimensional system
of size $L$ obeys the scaling form $\langle M_B \rangle \sim
L^{d_B}G(r/L)$, where the function $G(r/L)$ can be approximated by a
power-law for any value of $r/L$. Their results from numerical
simulations with {\it uncorrelated structures} indicate that, for the
case where $r\approx L$, the distribution of backbone mass is peaked
around $L^{d_B}$. When $r \ll L$, the distribution follows a power-law
behavior.

The fact that the geometry of real rocks is generally random does not
necessarily imply that their disordered morphology is spatially
uncorrelated. In other words, the probability for a small region 
in the system to have a given porosity (or permeability) may not be
independent of the probability associated with other locations. 
This is certainly true for some types of rocks and geological fields 
where geometrical and/or transport properties can be adequately 
characterized in terms of their spatial correlations \cite{Sahimi95}. 
It is under this framework that the correlated percolation model 
\cite{Havlin88,Havlin89,Prakash92,Makse96} represents a more 
realistic description for the pore space in terms of structure and
transport. In a previous work \cite{Makse00}, the hydrodynamic
dispersion behavior of percolation porous media with spatial
correlations has been investigated. More recently, we studied the
displacement dynamics between two fluids flowing through correlated
percolation clusters \cite{Araujo02}. It was found that the presence
of correlations can substantially modify the scaling behavior of the
distributions of relevant transport properties and, therefore, their
universality class. In the present paper we extend the previous work
of Barth\'el\'emy {\it et al.} \cite{Barth99} to investigate the
effect of spatial long-range correlations on the statistics of the
backbone mass connecting two ``wells'' in a two-dimensional
percolation geometry. In Sec.~II, we present the mathematical model to
simulate long-range spatial correlations. The results and analysis of
the numerical simulations are shown in Sec.~III, while the conclusions
are presented in Sec.~IV.

\section{Model}

Our model for the porous medium is based on a two-dimensional site
percolation cluster of size $L$ at criticality \cite{Stauffer94} 
where correlations among the elementary units of the lattice are
systematically introduced \cite{Prakash92,Makse96}. For a given
realization of the correlated network, we extract the percolation
backbone between two sites $A$ and $B$ separated by an Euclidian
distance $r \ll L$ that belong to the infinite cluster and are
sufficiently far from the edges of the system to prevent boundary
effects. The correlations are induced by means of the Fourier
filtering method, where a set of random variables $u(\textbf{r})$ is
introduced following a power-law correlation function of the form
\begin{equation}
\langle u(\textbf{r})u(\textbf{r}+\textbf{R})\rangle \propto
R^{-\gamma} \qquad\qquad [0<\gamma\leqslant 2]~.
\label{correl}
\end{equation}
Here $\gamma=2$ is the uncorrelated case and $\gamma\approx 0$
corresponds to the maximum correlation. The correlated variables
$u(\textbf{r})$ are used to define the occupancy $\zeta (\textbf{r})$
of the sites
\begin{equation}
\zeta (\textbf{r})=\Theta [\phi -u(\textbf{r})]~,
\label{zeta}
\end{equation}
where $\Theta$ is the Heavyside function and the parameter $\phi$ is
chosen to produce a network at the percolation threshold. Due to
computational limitations, we restricted our simulations to two values
of $\gamma=2$ and $0.5$, corresponding to uncorrelated and correlated
percolation structures, respectively. For each value of $\gamma$, we
performed simulations for $100~000$ network realizations of size $L
\times L$, where $L=1024$, and different values of the ``well''
distance $r$ to compute the distribution $P(M_B)$ and the cumulative
distribution $F(M_B)$ of backbone mass 
\begin{equation}
F(M_B)=\int_{M_{B}}^{\infty}P(M)dM~.
\label{cumul_scaling}
\end{equation}

\section{Results}

In Fig.~1 we show the log-log plot of typical cumulative distributions
of backbone mass $F(M_B)$ for uncorrelated as well as correlated
morphologies. It is clear from this figure that $F(M_B)$ displays
power-law behavior for intermediate mass values in both cases. In
addition, the scaling region is followed by a sudden cutoff that
decays faster than exponential. A similar behavior has been
observed experimentally \cite{Meibom96} and through numerical
simulations \cite{Inaoka97,Diehl00,Astrom01,Diehl01} for the
phenomenon of {\it impact fragmentation}. Based on these features, 
we argue that $F(M_B)$ should obey the following scaling ansatz:
\begin{equation}
F(M_B) \sim M_B^{-\alpha}\exp{\left[-\left(\frac{M_B}{M_0}
\right)^\eta\right ] }~,
\label{cumul}
\end{equation}
where $\alpha$ is a scaling exponent and $M_0$ and $\eta$ are 
cutoff parameters. For comparison with the approach presented in
Ref.~\cite{Barth99}, here we determine these parameters from the
simulation results in terms of their corresponding distributions of
backbone mass $P(M_B)$. From Eq.~(\ref{cumul}), it follows that
\begin{equation}
P(M_B) \sim {M_B}^{-(\alpha+1)}f(M_B/M_0)~,
\label{distr}
\end{equation}
where the function $f(M_B/M_0)$ has the form
\begin{equation}
f(M_B/M_0)=\left[\alpha+\eta\left(\frac{M_B}{M_0}\right)^{\eta}\right]
\exp\left[-\left(\frac{M_B}{M_0}\right)^{\eta}\right]~.
\label{func}
\end{equation}
Figure~2 shows the distributions $P(M_B)$ generated for the
uncorrelated case ($\gamma=2$) and different values of $r$. They all
display a lower cutoff of order $r$ (the smallest backbone possible is
a straight line connecting points $A$ and $B$) and an upper cutoff of
order $L^{D_B}$, where a ``bump'' can also be observed \cite{Barth99}.
For comparison, each of these distributions has been rescaled by
its corresponding value at the position of this bump. The smaller
the distance $r$ is between the wells, larger is the range over which
the scaling term of Eq.~(\ref{distr}) holds, $P(M_B) \sim
{M_B}^{-(\alpha+1)}$. The solid line in Fig.~2 corresponds to the best
nonlinear fit we found for the $r=8$ data in both the scaling and
cutoff zones with $\alpha=0.255$, $M_0=3.32 \times 10^4$ and
$\eta=1.5$. Because the sampling of the backbone mass should be
equivalent to the sampling of blobs in the percolation cluster
\cite{Stanley77}, it is possible to draw a direct relation between the
scaling exponent $\alpha$ and the fractal dimension of the backbone
\cite{Herrmann84}. Accordingly, the exponent $\tau$ governing the blob
size distribution can be calculated as $\tau=d/d_{B}+1$. Since the
exponent $\tau_{B}=\alpha+1$ governs the statistics of the entire
backbone, we obtain by integration that $\alpha=d/d_B-1$,
which gives a fractal dimension of $d_B \approx 1.6$. This is in good
agreement with the current numerical estimate of $d_B=1.6432\pm
0.0008$ \cite{Grass99}.

We now turn to the case with spatial long-range correlations. As shown
in Fig.~3, Eq.~(\ref{distr}) (solid line) also fits very well the
simulation data for $P(M_{B})$ with $\gamma=0.5$ and $r=16$, both in
the scaling region and in the cutoff zone. The parameter set obtained
with a nonlinear estimation algorithm includes $\alpha=0.075$,
$M_0=5.4 \times 10^5$, and $\eta=2$. Compared to the uncorrelated
geometry, the results shown in Fig.~3 for large masses clearly
indicate the presence of a narrow plateau followed by a much more
pronounced bump and a sharper cutoff profile. These features are
consistent with the exponent $\alpha \approx 0$ and also reflect the
fact that $\eta$ is significantly larger for the correlated case. The
resulting fractal dimension of the backbone $d_B \approx 1.86$ is in
good agreement with previous estimates for correlated structures
generated with $\gamma=0.5$
\cite{Prakash92,Araujo02}.

For completeness, we show that the scaling ansatz used to fit the
simulation results for $P(M_B)$ is consistent with the observed
behavior for $F(M_B)$ over the whole range of relevant backbone
masses. Precisely, the solid lines in Fig.~1 correspond to
Eq.~(\ref{cumul_scaling}) with the same set of parameters used to fit
the simulation data for $\gamma=0.5$ and $2$ in terms of
Eq.~(\ref{distr}). Although close to zero, the value of
$\alpha=0.075$ for the correlated geometry is sufficiently large to
characterize the power-law signature of the cumulative mass
distribution $F(M_B)$. The differences in the exponent $\alpha$
obtained for correlated and uncorrelated cases can be explained in
terms of the morphology of the conducting backbone. As $\gamma$
decreases, the backbone becomes gradually more compact
\cite{Prakash92}. This distinctive aspect of the correlated geometry
tends to increase the value of $d_B$ and therefore reduce the value of
$\alpha$ as the strength of the long-range correlations increases
(i.e., $\gamma$ decreases).  As in Ref.~\cite{Barth99}, here we also
investigate the scaling behavior of the average backbone mass $\langle
M_B \rangle$, but for a correlated geometry. In Fig.~4 we show the
log-log plot of $\langle M_B \rangle$ against $r$ for $\gamma=0.5$ and
three different values of $L$. The approximation proposed in
Ref.~\cite{Barth99} for uncorrelated networks when $r \ll L$ reveals
that $\langle M_B \rangle$ should scale as
\begin{equation}
\langle M \rangle \sim L^{d_{B}-\psi}r^{\psi}~,
\label{average}
\end{equation}
where the exponent $\psi$ is the codimension of the fractal backbone,
i.e. $\psi=d-d_B$. Using Eq.~(\ref{average}) and $d_B=1.85$, the inset
of Figs.~4 shows the data collapse obtained by rescaling $M_B$ and $r$
to the corresponding values of $L^{d_B}$ and $L$, respectively. From
the least-squares fit to the data in the scaling region, we obtain the
exponent $\psi=0.17 \pm 0.03$. This result is in good agreement with
the estimated value $0.15$ for the fractal codimension. 

Finally, it is also interesting to investigate the case where
$r\approx L$. As for uncorrelated clusters \cite{Barth99}, we expect the
distribution $P(M_B)$ for correlated geometries to obey the simple
scaling form
\begin{equation}
P(M_B) \sim \frac{1}{L^{d_B}}g\left(\frac{M_B}{L^{d_B}}\right)~,
\label{large_r}
\end{equation}
where $g$ is a scaling function. In Fig.~5 we show results from
simulations with correlated networks generated with $\gamma=0.5$ for
$r\approx L$ and three different values of $L$. It can be seen that
the probability distribution is peaked around an average value
$\langle M_B \rangle$ of the order of $L^{d_B}$. As depicted,
Eq.~(\ref{large_r}) can indeed be used to collapse the data. The value
adopted for the fractal dimension of the backbone, $d_B=1.86$, 
is compatible with the corresponding degree of correlation imposed.

\section{Summary}

We have studied the scaling characteristics of the backbone mass
distribution between two sites in two-dimensional percolation lattices
subjected to long-range correlations. A scaling ansatz that is capable
of describing the power-law region as well as the complex details of the
cutoff profile is proposed and it is shown to be applicable for both
correlated and uncorrelated structures. Based on the results of
extensive simulations, we find that the presence of long-range
correlations can substantially modify the scaling exponents of these
distributions and, therefore, their universality class. We explain
this change of behavior in terms of the morphological differences
among uncorrelated and correlated pore spaces generated at
criticality. Compared to the uncorrelated case, the backbone clusters
with spatial long-range correlations have a more compact geometry. The
level of compactness depends, of course, on the degree of correlations
$\gamma$ introduced during the generation process.

\section{Acknowledgments}

This work has been supported by CNPq, CAPES and FUNCAP.

\newpage

\begin{figure}
\includegraphics[width=8.0cm]{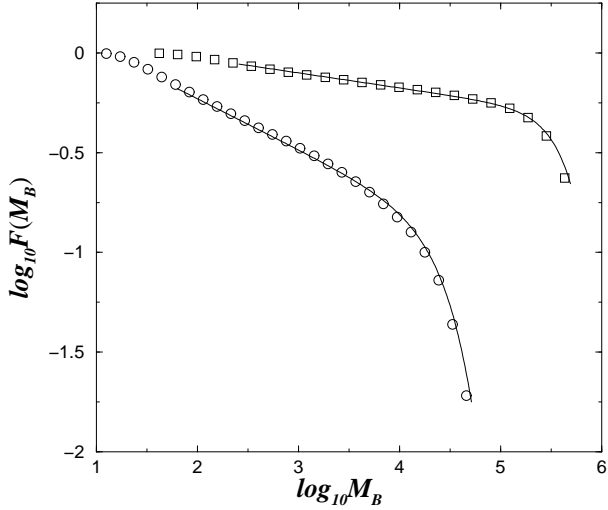}
\caption{Log-log plot of the cumulative distribution of backbone
mass for uncorrelated ($\gamma=2$, circles), and correlated networks
($\gamma=0.5$, squares). The solid lines correspond to the scaling
function Eq.~(\ref{cumul}) with the same values of the parameters
obtained from the best fit to the data of Eq.~(\ref{distr}).}
\end{figure}

\begin{figure}
\includegraphics[width=8.0cm]{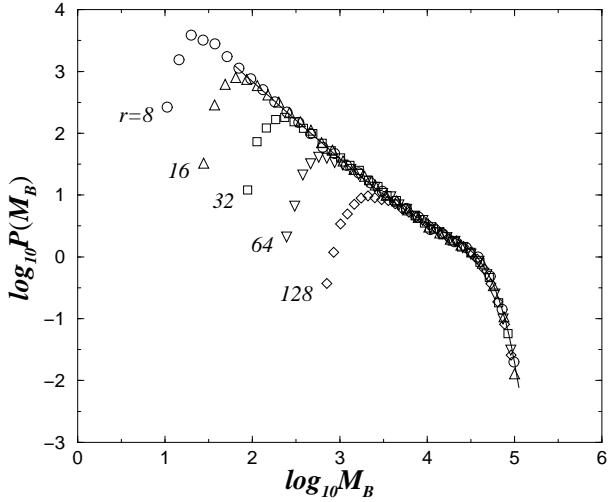}
\caption{Log-log plot of the distribution of backbone mass of
uncorrelated percolation networks for $L=1024$ and $r=8$ (circles),
$16$ (triangles up), $32$ (squares), $64$ (triangles down), and $128$
(diamonds). The solid line is the best fit of Eq.~(\ref{distr})
to the data for the scaling region and lower cutoff, with parameters
$\alpha=0.255$, $\eta=1.5$, and $M_0=3.32 \times 10^4$.}
\end{figure}

\begin{figure}
\includegraphics[width=8.0cm]{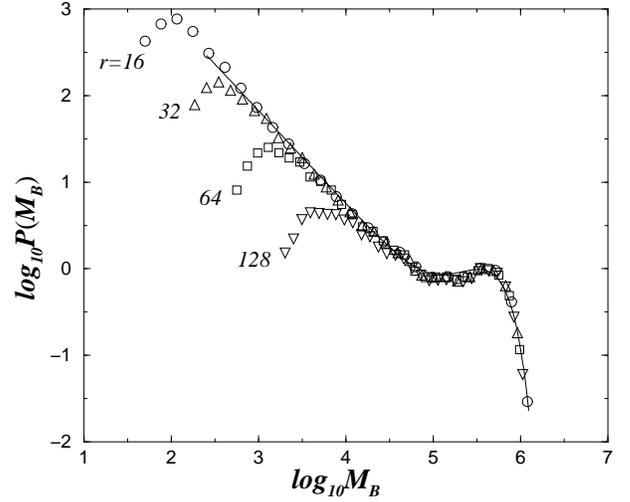}
\caption{Log-log plot of the distribution of backbone mass of
correlated percolation networks for $L=1024$ and $r=16$ (circles),
$32$ (triangles up), $64$ (squares), and $128$ (triangles down). The
solid line is the best fit of Eq.~(\ref{distr}) to the data for the
scaling region and lower cutoff, with parameters $\alpha=0.075$,
$\eta=2.0$, and $M_0=5.4 \times 10^5$. The degree of correlation is
$\gamma=0.5$.}
\end{figure}

\begin{figure}
\includegraphics[width=8.0cm]{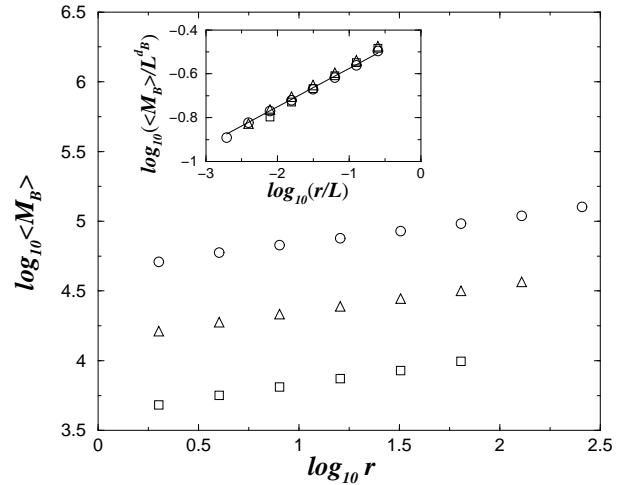}
\caption{Log-log plot of the average backbone mass $\langle M_{B} \rangle$
for correlated structures ($\gamma=0.5$ against the distance $r$ for
$L=$ (circles), $$ (triangles) and $$ (squares). The inset shows the
best data collapse obtained by rescaling $M_B$ and $r$ to $L^{1.85}$ and
$L$, respectively. The least-squares fit to the data gives the scaling
exponent $\psi=0.17 \pm 0.03$.}
\end{figure}

\begin{figure}
\includegraphics[width=8.0cm]{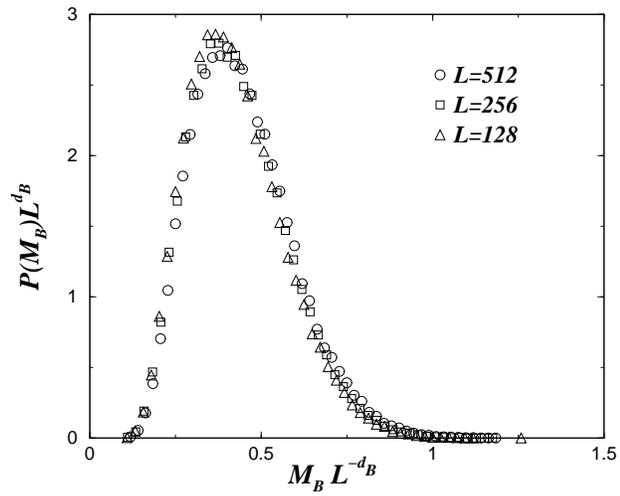}
\caption{Data collapse of the distribution of backbone mass for
correlated networks ($\gamma=0.5$) and three values of $r \approx L$.
For each value of $L$, $100~000$ realizations have been generated to
produce the statistics. Equation~(\ref{large_r}) with $d_B=1.86$ has
been used to collapse the results.}
\end{figure}

\end{multicols}

\end{document}